\documentclass[a4paper]{jpconf}
\usepackage{graphicx}
\usepackage{amsmath}
\usepackage{epstopdf}
\begin{document}

\title{Heavy quark dynamics within a Boltzmann transport model: radiative vs collisional energy loss}

\author{Gabriele Coci$^{1,2}$, Francesco Scardina$^2$ and Vincenzo Greco$^{1,2}$}

\address{$^1$ Department of Physics and Astronomy, University of Catania, Via S. Sofia 64, ZIP-95125 Catania (Italy)}
\address{$^2$ Laboratori Nazionali del Sud, INFN-LNS, Via S. Sofia 62, ZIP-95125 Catania (Italy)}

\ead{coci@lns.infn.it}

\begin{abstract}
We discuss the propagation of heavy quarks (charm and bottom) through the QGP by means of a relativistic Boltzmann transport approach including both collisional and radiative energy loss mechanisms.
In particular we investigate the impact of induced gluon radiation by dynamical QCD medium implementing in our transport model a formula for the emitted gluon spectrum calculated in a higher-twist scheme. 
We notice that in the region of high transverse momentum ($p_T > 10$ GeV) radiative processes play an essential role giving a dominant contribution to the generation of $R_{AA}$ and $v_2$ at momentum values for which the energy loss by collisions is in the perturbative regime.
\end{abstract}
\section{Introduction}
Heavy quarks (HQs) represent an ideal probe for investiganting the properties of hot and dense QCD matter created at ultra-Relativistic Heavy Ion Collisions (uRHICs). 
They are created at early stages of collisions and due to their large masses they have a large thermalization time with respect to the bulk medium. Thus they conserve memory of propagation during the entire space-time evolution of the QGP. 
As they diffuse through the medium, HQs suffer collisions and radiative effects as a kind of gluon bremsstrahlung. Both processes are responsible for the substantial quenching of high-$p_T$ HQs
and also for the coupling of their dynamics with the bulk collective flow.
The two key observables in HQ sector are the nuclear suppression factor $R_{AA}(p_T)$, i.e. the ratio between the spectra of heavy flavor hadrons and their decay products measured in nucleus-nucleus collisions with the same spectra in proton-proton ones, and the elliptic flow $v_2(p_T)$ which is a measure of the developed anisotropies in the angular distributions in momentum space. 
Both observables have been measured at RHIC and LHC facilities \cite{PHENIX_Adare2011_HQ} \cite{ALICE_Abelev2012_HQ} and the challenge of each theoretical framework is to give a consistent explanation of the surprising low $R_{AA}$ and high $v_2$ observed in the HQ sector, hence solving the so called $R_{AA}-v_2$. 

\section{Our transport model for heavy quarks in QGP}
We describe the HQ propagation inside the QGP by means of relativistic Boltzmann equation which is written as follows:
\begin{equation}\label{eq:HQBoltzmann}
p^{\mu}\partial_{\mu}f_{HQ}(x,p) = C[f_{HQ}](x,p)
\end{equation}
\noindent $f_{HQ}$ is the HQ phase-space distribution function and $C[f_{HQ}]$ is the relativistic Boltzmann-like collision integral which contains all the interaction mechanisms between HQs and light partons of the bulk, which for simplicity we assume to be formed by gluons.
We consider only elastic two-body collisions whose total cross section is derived from the scattering matrices of $g+HQ \rightarrow g+HQ$ processes calculated at pQCD leading-order \cite{Combridge79}. A realistic Debye screening mass $m^2_D = 4\pi \alpha_s T^2$ for regulating the {\it t}-channel gluon propagator and the pQCD running of the coupling constant $\alpha_s(T)$ with bulk temperature {\it T} are also included. 
The integrated distribution function for the medium components contained in $C[f_{HQ}]$ is determined in analogous manner by solving the related Boltzmann-Vlasov equation \cite{ctQGP_PRC89}. The Boltzmann equation is solved numerically 
dividing the coordinate space in a 3D lattice and the test particle method is used to sample the distribution function in each cell. The collision integral is solved by means of the stochastic algorithm for evaluating the collision probability \cite{XuGreiner2005} \cite{ctQGP_PLB724}. \\
\noindent The goal of this work is to include inside this framework a model which could account for the energy loss of HQs by inelastic processes. With this aim we implemented a mechanism for medium-induced gluon radiation calculated in multiple-scattering approximation and pQCD factorization scheme which was already included in other models where HQ dynamics was described by Langevin \cite{CaoQinBassMuller_FKL_2013} or Linearized Boltzmann equation \cite{CaoLuoQinWang_LBT_2016}. The differential spectrum of emitted gluons is derived from the higher-twist calculation for the energy loss of fast partons \cite{WangGuo_NuclPhysA_2001} 
\cite{Majumder_PRD85_2012} and heavy quarks \cite{ZhangWangWangXN_NuclPhysA757_2005} and it is given by the following expression

\begin{equation}\label{eq:DNg}
\frac{dN_g}{dxdk_{\perp}^2dt} =  \frac{2 \alpha_s C_A P(x)}{\pi k_{\perp}^4}\hat{q}\left( \frac{k_{\perp}^2}{k_{\perp}^2+x^2M^2} \right)^4 \sin^2\left( \frac{t}{2\tau_f}\right)
\end{equation}

\noindent where $x$ and $k_{\perp}$ are respectively the fraction of HQ energy $E$ carried away by the emitted gluon and the transverse gluon momentum with respect to the propagating direction.\\
$P(x)$ is the gluon splitting function derived from medium-modified DGLAP equations \cite{DengWangXN_PRC81_2010}, while the $\hat{q}$ represents the HQ transport coefficient depending on  {\it E} and {\it T}. In this initial work we simply relate the $\hat{q}$ with the momentum-space diffusion coefficient, i.e. $\hat{q}=4\kappa$, which has a better significance if Fokker-Planck approach for HQ dynamics is applied. In further study a more precise treatment of $\hat{q}$ within our Boltzmann transport model will be considered referring also to the Jet Collaboration efforts done in extracting such jet quenching parameter \cite{JetCollaboration_Burke2014}.
Destructive interference known as LPM effect \cite{LPM_original} which occurs if the time between two successive collisions is smaller than the formation time $\tau_f = 2Ex(1-x)/(k_{\perp}^2+x^2M^2)$ and suppression of radiation at small angle of emission (''dead-cone'' effect \cite{Dokshitzer_Kharzeev_PLB519_2011}) due to non-negligible HQ mass $M$ are both included as visible from Eq.~\eqref{eq:DNg}. \\
The differential spectrum is numerically integrated in $x$ and $k_{\perp}^2$ and the result multiplied by the timestep of evolution $\Delta t$ in order to get the single-emission radiation probability $P_{rad}(t,T,E)$. 
A physical lower cut-off of the gluon energy equivalent to $m_D$ is considered in order to avoid the divergence at $x \rightarrow 0$. 
In the same manner we did for the collision probability, we use $P_{rad}(t,T,E)$ to sample the number of test particles which radiate at each timestep and we distribute the four-momentum components of the emitted gluon according to the differential spectrum Eq.~\eqref{eq:DNg}.
By such a fully reconstruction of the final state we are able to relax the path-length approximation for HQ propagation and include its diffusion during the radiation mechanism, which we predict has a remarkable effect on $v_2$ formation (see Fig.~\ref{fig:v2_rhic_fixed}) since it couples the HQ dynamics at high $p_T$ more strongly to the collective flow of the bulk, keeping a similar suppression pattern.

\section{Results at RHIC} 
We have carried out simulations at RHIC (Au+Au at $\sqrt{s}=200$ AGeV, $b=7.5$ fm) events. The initial conditions in the coordinate space are given by the standard geometrical Glauber model, while for initial distributions in momentum space we employ a Boltzmann-J{\"u}ttner function for bulk gluons and a power law function calculated within pQCD NLO \cite{CacciariNasonVogt} for charm quarks.\\
Our purpose in this study was to investigate the impact of various energy loss scenarios on $R_{AA}$ and $v_2$ observables, therefore we have not included the hadronization mechanisms yet. \\
\noindent It is well know that pQCD alone is not able to reproduce the quenching in HQ sector. The problem is usually cleared by introducing a constant {\it k} factor to enhance cross section for HQ scatterings. The physical reason for introducing such factor is the idea to include non-perturbative effects inside the HQ dynamics which become dominant at low $p_T$ \cite{HeesMannarelliRappGreco2008}. However in the region of high $p_T$ where radiative energy loss is much higher than collisional one, such non-perturbative effects should be negligible. Referring to \cite{CaoLuoQinWang_LBT_2016} we model this effect by multiplying both collisional and radiative probabilities with a $p_T$-dependent {\it k}-enhancement which for simplicity is assumed to have a gaussian profile:
$k(p_T) = \left[ 1 + \left( k_0 -1 \right)\exp{\left(-\frac{p_T^2}{2 \sigma_p^2}\right)} \right].$

\begin{figure}[!h]
\begin{minipage}{18pc}
\includegraphics[width=\textwidth ,height=0.75\textwidth]{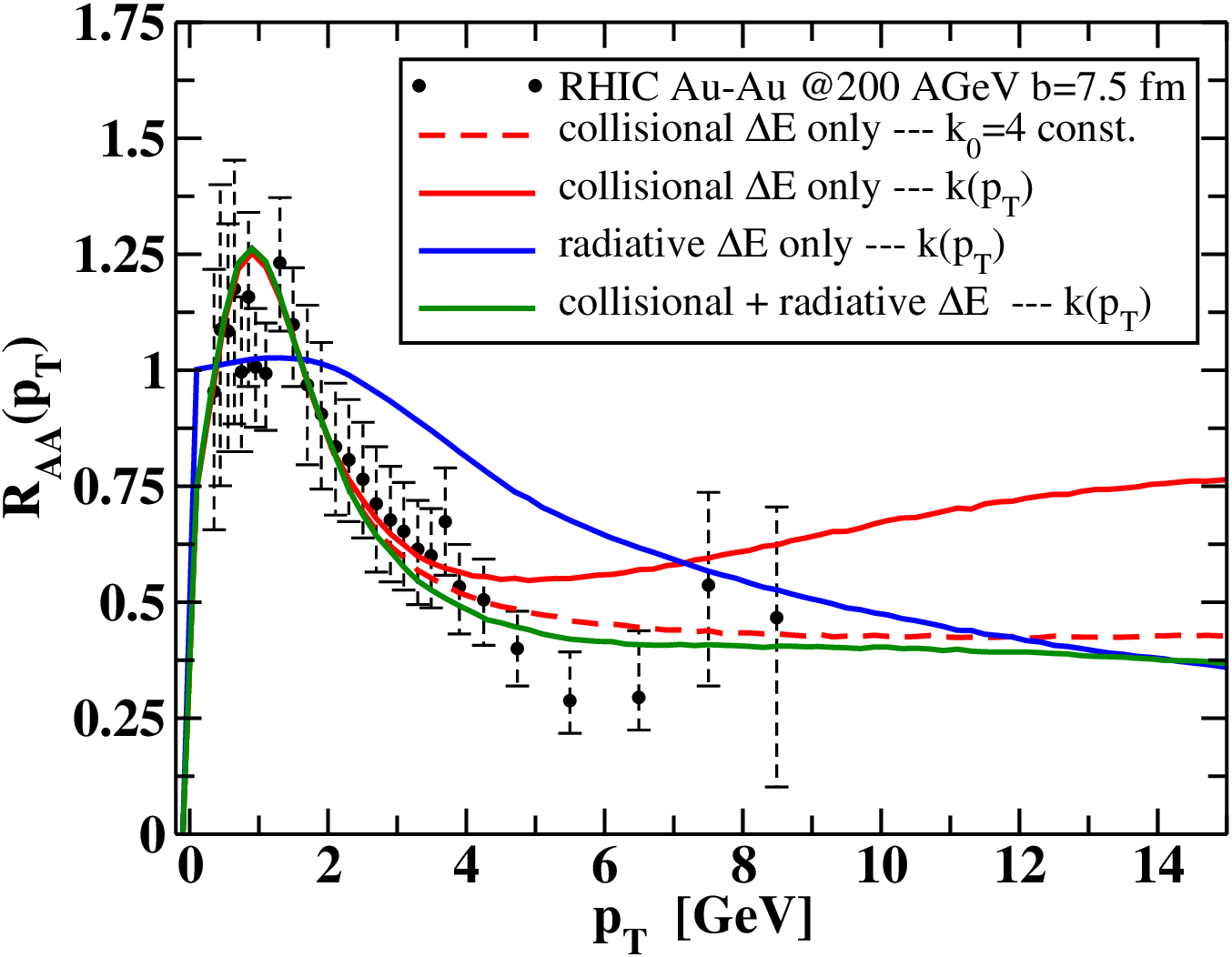}
\caption{\label{fig:RAA_rhic1} Measured $R_{AA}$ at RHIC for single-non photonic electrons compared to the suppression of charm quarks. The red dashed line is the non-physical collisional scenario with constant $k_0$=4, while the solid lines are respectively collisional only (red), radiative only (blue) and collisional+radiative scenarios (green) with $k_0$=4 and $\sigma_p$=5 GeV.}
\end{minipage}\hspace{2pc}%
\begin{minipage}{18pc}
\includegraphics[width=\textwidth ,height=0.735\textwidth]{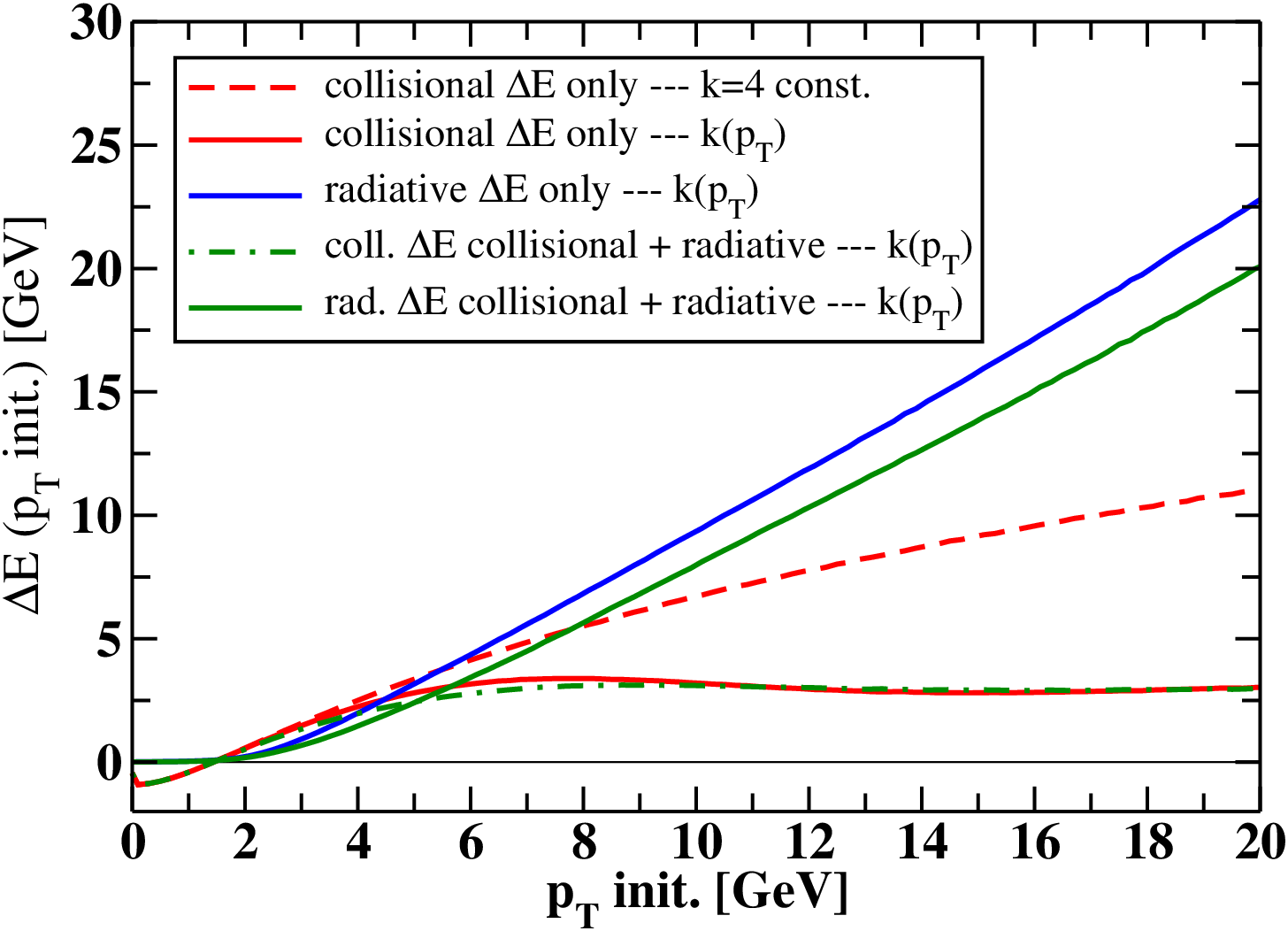}
\caption{\label{fig:DE_rhic1} Collisional and radiative energy loss distributions for charm quarks as function of initial $p_T$. Colours and parameters are the identical to the $R_{AA}$ plot in Fig.~\ref{fig:RAA_rhic1}. The green dashed and green solid lines are respectively the collisional and radiative contributions in case both energy loss processes are considered.}
\end{minipage} 
\end{figure}

\noindent We set the width parameter $\sigma_p$ in order to reach the condition $k(p_T) \rightarrow 1$ at $p_T > 10$ GeV with a good approximation. The amplitude $k_0$ which corresponds to the maximum enhancement in the limit $p_T \rightarrow 0$ is fixed instead by fitting the $R_{AA}$ with a non-physical scenario where only collisions occur. The $R_{AA}$ suppression in all studied scenarios is shown in Fig.~\ref{fig:RAA_rhic1}. In Fig.~\ref{fig:DE_rhic1} the energy loss distributions of charm quarks as function of their initial $p_T$ is plotted.\\

\begin{figure}[!h]
\begin{minipage}{18pc}
\includegraphics[width=\textwidth, height=0.75\textwidth]{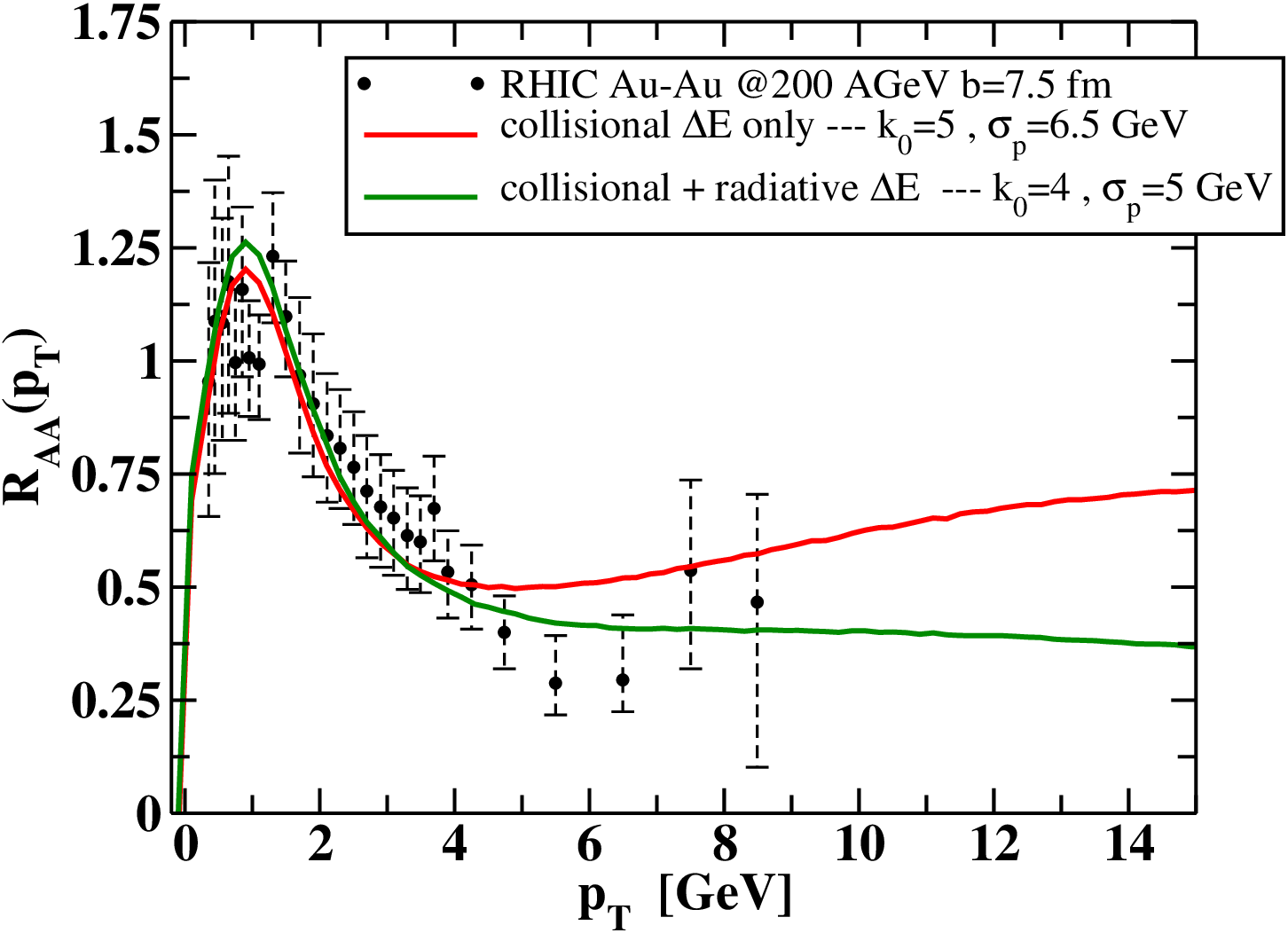}
\caption{\label{fig:RAA_rhic_fixed} Suppression factor $R_{AA}$ obtained at RHIC energies fitted by pure collisional quenching (red line) and collisional+radiative energy loss scenario (green line). Values of $k_0$ and $\sigma_p$ parameters are reported in the legend.}
\end{minipage}\hspace{2pc}%
\begin{minipage}{18pc}
\includegraphics[width=\textwidth, height=0.745\textwidth]{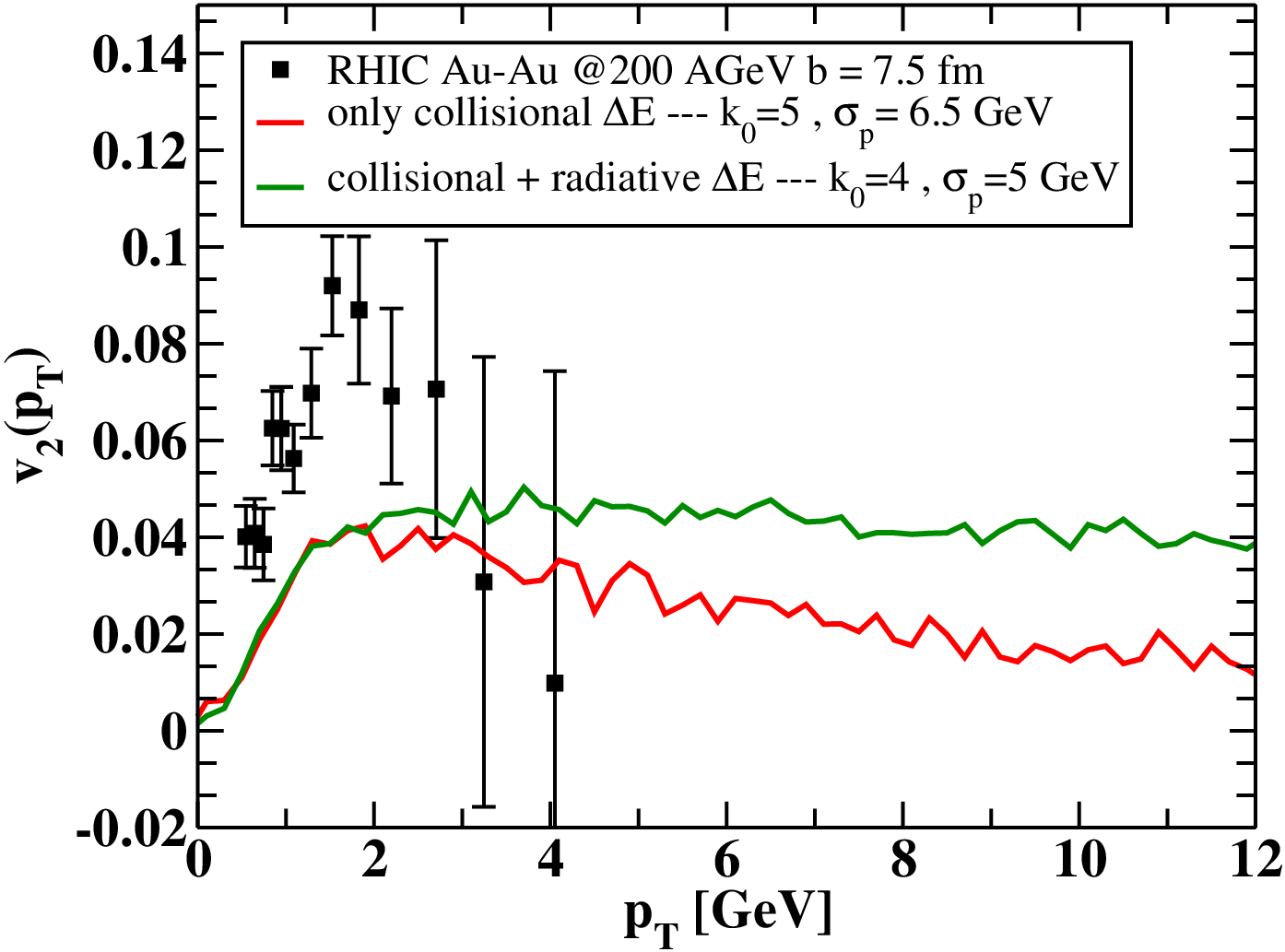}
\caption{\label{fig:v2_rhic_fixed} Comparison between elliptic flow $v_2$ measured at RHIC energies produced in a scenario where only collisions are considered (red line) against a more complete case where gluon radiation is also included (green line).}
\end{minipage} 
\end{figure}

\noindent Once the $R_{AA}$ is fixed at least in the region of experimental points by tuning $k_0$ and $\sigma_p$, we can look at the impact on the elliptic flow as it is visible from Fig.~\ref{fig:v2_rhic_fixed}. The production of $v_2$ at  $p_T > 10$ GeV is mostly determined by the radiative mechanism. As said before, we notice that in our approach the formation of $v_2$ is naturally beyond the simple path-length approximation; in fact the HQs can both radiate or absorb energy and can deviate from their trajectory due to gluon emission. 
The discrepancy with experimental data at $p_T < 2-3$ GeV is known to be reduced by means of a {\it T}-dependent {\it k}-factor in the collisional energy loss \cite{ctQGP_PLB747}.

\section{Conclusions}
In this work we studied the HQ propagation in QGP by means of Boltzmann equation and we implemented both a radiative and collisional energy loss mechanism. 
For the low $p_T$ region we modelled non-perturbative effects using a $k(p_T)$ enhancement which disappears once the momentum scale is sufficiently higher to have a perturbative regime ($p_T \approx 10$ GeV). 
In future, we will concentrate on the possibility of implementing a multiple-emission algorithm in order to increase the role of coherent effects during radiative processes and also include the emitted gluons within the bulk dynamics and the subsequent rescattering.
We will focus also on accounting non-perturbative effects with a different {\it T}-dependent behaviour of the interaction strenght in order to study the $v_2$ formation at low $p_T$.
The goal is to include the HQ transport model in a more complete description of uRHICs where also initial-stage dynamics and hadronization mechanisms are accounted.  
The aim is not only to solve the $R_{AA}-v_2$ puzzle but also to investigate more differential observables, such as azimuthal correlations, which are going to be measured both at RHIC and LHC energies hopefully.


\section*{References}
\begin{thebibliography}{9}

\bibitem{PHENIX_Adare2011_HQ} PHENIX Collaboration, A. Adare {\it et al.}, {\it Phys. Rev.} C {\bf 84}, 044905 (2011), arXiv:1005.1627

\bibitem{ALICE_Abelev2012_HQ} ALICE Collaboration, B. Abelev {\it et al.}, {\it J. High Energy Phys.} 09, 112 (2012).
 

\bibitem{Combridge79} B. Combridge, {\it Nucl. Phys.} B vol. {\bf 151}, pp. 429-456 (1979).

\bibitem{ctQGP_PRC89} M. Ruggieri, F. Scardina, S. Plumari and V. Greco, {\it Phys. Rev.} C {\bf 89}, 054914 (2014).

\bibitem{XuGreiner2005} Z. Xu, C. Greiner, {\it Phys. Rev.} C {\bf 71}, 064901 (2005).

\bibitem{ctQGP_PLB724} F. Scardina, M. Colonna, S. Plumari and V. Greco, {\it Phys. Lett.} B {\bf 724}, pp. 296-300 (2013).




\bibitem{CaoQinBassMuller_FKL_2013} S. Cao, G.-Y. Qin, S. Bass and B. M{\"u}ller, {\it Nucl. Phys.} A {\bf 904-5}, pp. 653c-656c (2014). 

\bibitem{CaoLuoQinWang_LBT_2016} S. Cao, T. Luo, G.-Y. Qin and X.-N. Wang, {\it Phys. Rev.} C {\bf 94}, 014909 (2016). 

\bibitem{WangGuo_NuclPhysA_2001}  X.-N. Wang, X.-F. Guo, {\it Nucl. Phys.} A {\bf 696}, 788 (2001).


\bibitem{Majumder_PRD85_2012} A. Majumder, {\it Phys. Rev.} D {\bf 85}, 014023 (2012).

\bibitem{ZhangWangWangXN_NuclPhysA757_2005} B.-W. Zhang, E. Wang and X.-N. Wang, {\it Nucl. Phys.} A {\bf 757}, pp. 493-524 (2005).


\bibitem{DengWangXN_PRC81_2010} W.-T. Deng and X.-N. Wang, {\it Phys. Rev.} C {\bf 81} 024902 (2010).

\bibitem{JetCollaboration_Burke2014} Jet Collaboration, K.M. Burke {\it et al.}, {\it Phys. Rev.} C {\bf} 90, 014909 (2014).


\bibitem{LPM_original} Landau and Pomeranchuk, {\it Dokl. Akad. Nauk Ser. Fiz} {\bf 92} 535 (1953); Migdal, {\it Phys. Rev.} {\bf 103}, 1811 (1956).

\bibitem{Dokshitzer_Kharzeev_PLB519_2011} Y.L. Dokshitzer and D.E. Kharzeev, {\it Phys. Lett.} B {\bf 519} 199 (2001).

\bibitem{CacciariNasonVogt} M. Cacciari, P. Nason and R. Vogt, {\it Phys. Rev. Lett.} {\bf 95}, 122001 (2005).

\bibitem{HeesMannarelliRappGreco2008} H. van Hees, M. Mannarelli, V. Greco and R. Rapp, {\it Phys. Rev. Lett.} 100, 192301 (2008).

\bibitem{ctQGP_PLB747} S.K. Das, F. Scardina, S. Plumari and V. Greco, {\it Phys. Lett.} B {\bf 747}, 260-264 (2015).


\end{thebibliography}
\end{document}